\documentclass[]{an}
\usepackage{graphicx}
\usepackage{times}
\Volume{00}              
\Year{0000}              
\Month{00}               
\Pagespan{000}{000}      

\begin{document}
\lhead[\thepage]{I.~Negueruela: Luminous Be stars}
\rhead[Astron. Nachr./AN~{\bf XXX} (200X) X]{\thepage}
\headnote{Astron. Nachr./AN {\bf 32X} (200X) X, XXX--XXX}

\title{A Search for Luminous Be stars\fnmsep\thanks{Based on
observations obtained at the the Isaac Newton Telescope (La Palma,
Spain) and Observatoire de Haute Provence (CNRS, France)}} 

\author{I.~Negueruela}
\institute{Dpto. de F\'{\i}sica, Ing. de Sistemas y Teor\'{\i}a de
la Se\~{n}al, Universidad de Alicante, Apdo. 99, E03080 Alicante,
Spain}
\date{Received {date will be inserted by the editor}; 
accepted {date will be inserted by the editor}} 

\abstract{{As Be stars are restricted to luminosity classes III-V, but
early B-type stars are believed to evolve into supergiants, it is to
be expected that the Be phenomenon disappears at some point
in the evolution of a moderately massive star, before it reaches the
supergiant phase. As a first stage in an attempt to determine
the physical reasons of this cessation, a
search of the literature has provided a number of candidates to be Be
stars with luminosity classes Ib or II. Spectroscopy has been obtained
for candidates in a number of open clusters and associations, as well
as several other bright stars in those clusters. Among the objects
observed, HD~207329 is the best candidate to be a high-luminosity Be
star, as it appears like a fast-rotating supergiant with double-peaked
emission lines. The lines of HD~229059, in Berkeley~87, also appear
morphologically similar to those of Be stars, but there are reasons to
suspect that this object is an interacting binary. At slightly lower
luminosities, LS~I~$+56\degr$92 (B4\,II) and HD~333452 (O9\,II), also
appear as intrinsically luminous Be stars. Two Be stars in NGC~6913,
HD~229221 and HD~229239, appear to have rather higher intrinsic
magnitudes than their spectral type (B0.2\,III in both cases) would
indicate, being as luminous as luminosity class II objects in the same
cluster. HD~344863, in NGC~6823, is also a rather early Be star of
moderately high luminosity. The search shows that, though
high-luminosity Be stars do exist, they are scarce and, perhaps
surprisingly, tend to have early spectral types.} 
}  
\correspondence{ignacio@dfists.ua.es}

\maketitle

\section{Introduction}
\label{sec:intro}

In recent years, it has become obvious that rotation plays a
fundamental role in the evolution of massive stars (Maeder \& Meynet
2000). The initial rotational velocity determines the main-sequence
lifetime of a star, its post-main-sequence evolution and likely even
the nature of its final remnant after supernova explosion (Meynet \& Maeder
2000, 2003). However, one of the most obvious manifestations of stellar
rotation, the Be phenomenon, remains unexplained after more than a
century of study (see Porter \& Rivinius 2003 for a recent review).

The accepted definition of a Be star is that given by Collins (1987), namely, 
``a non-supergiant B star whose spectrum has, or had at some time, one or 
more Balmer lines in emission''. This definition may need some qualification,
as the Be phenomenon appears to extend into late O and early A subtypes, and
also because some objects (such as Herbig AeBe stars) are generally excluded 
from the class (see Porter \& Rivinius 2003 for a detailed discussion).
It is, however, widely accepted for the ``classical Be stars'', i.e., 
fast rotating moderately massive stars in which the emission lines are
produced in a circumstellar disk of material expelled from the photosphere.

\begin{figure}[ht]
\resizebox{\hsize}{!}
{\includegraphics[bb=60 165 520 600]{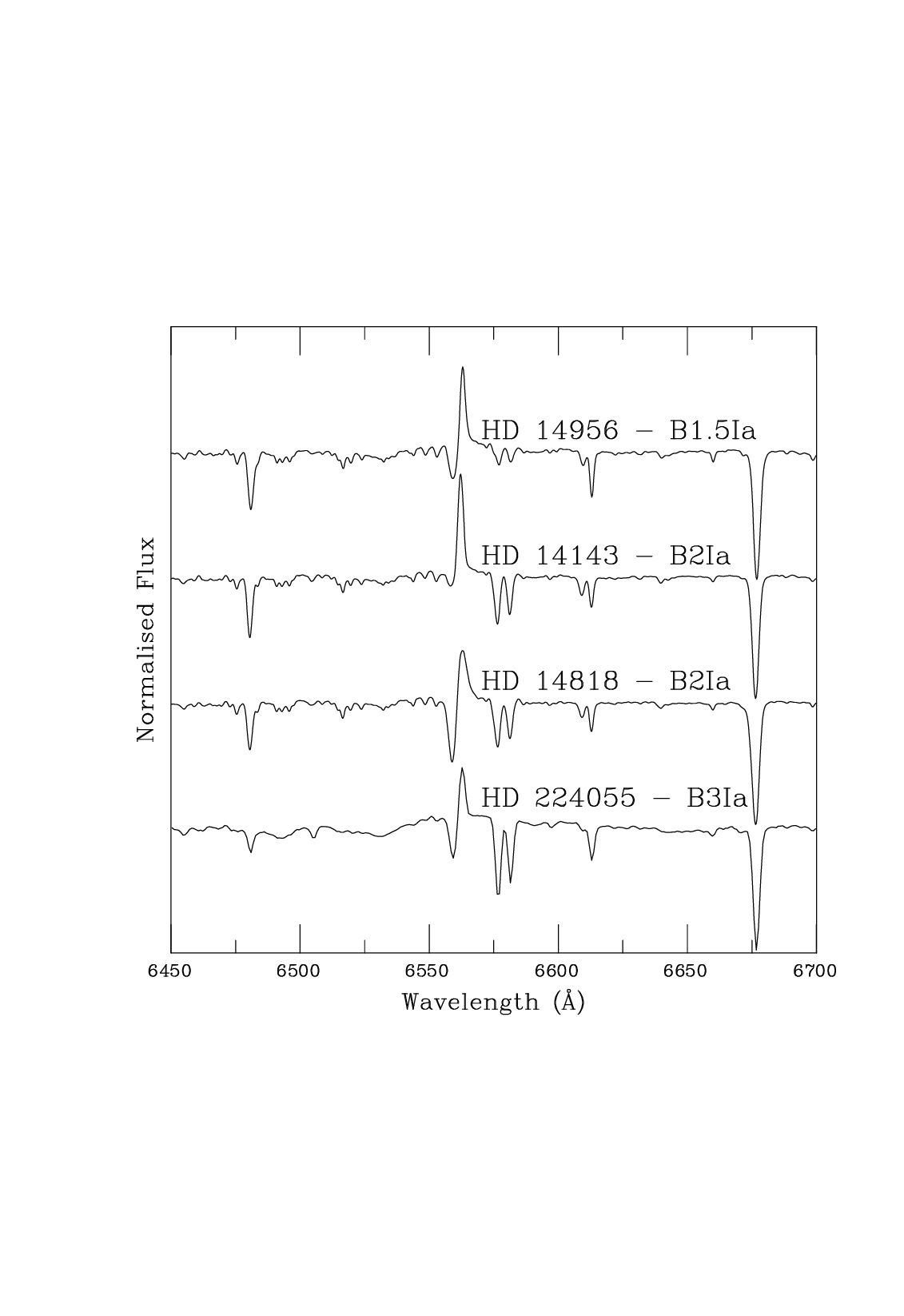}}
\caption{H$\alpha$ emission lines in four bright early-B
supergiants. The P-Cygni profiles may be considered typical of this
class of objects. Note that the absorption trough tends to disappear
towards early spectral types (it is very weak already in HD~14143),
making the profile look like a single peak, specially for stars B0 and
earlier. Note also the increase in the strength of the
\ion{C}{ii}~$\lambda\lambda$6578, 6583\AA\ doublet towards a maximum
at B3-4. Compare these profiles with those in Fig.~\ref{fig:minialpha}.}
\label{fig:superalpha}
\end{figure}

The ``non-supergiant'' part of the definition is generally understood to 
represent a warning against potential confusion. Many B-type supergiants
show relatively strong H$\alpha$ emission lines, which are not produced in
a circumstellar disk. These lines arise in the strong radiatively-driven
winds of the supergiants (e.g., Leitherer 1988) and are considered 
morphologically normal (i.e., not unusual for the spectral type). As a 
consequence of their different formation mechanisms, H$\alpha$
emission lines in classical Be stars and supergiants usually have
rather different shapes, with the lines in Be stars showing rotational
symmetry, while the lines in supergiants typically display P-Cygni
profiles (see Fig.~\ref{fig:superalpha}). 

There may, however, be a deeper interpretation to this qualification,
as it clearly states that no supergiant star displays the Be
phenomenon. As an important fraction of O9-B1 main sequence and giant
stars display the Be phenomenon (e.g., Zorec \& Briot 1997) and these 
stars are sufficiently massive to become blue supergiants at a later
stage, one would (naively) conclude that the Be phenomenon has to
{\it disappear} at some point during the post-Main-Sequence (post-MS) 
evolution of the star.

There are two immediately obvious physical effects to which this
cessation of the Be phenomenon could be linked. On the one hand, the Be
phenomenon is known to be related to fast rotation and supergiants do
not rotate fast. If the unknown cause of mass loss requires fast
rotation, then mass loss could stop as the rotation slows. Because of
angular momentum conservation, a star is expected to slow down as it
expands (cf. Steele 1999). On the
other hand, as the luminosity increases, the star develops a slow
radiative wind which could exert a force on the circumstellar disk
that would lead to its dissipation. The disk would be swept away by
radiation pressure even if suitable conditions for its formation were
still present.

In reality, things could not be so simple. The relevant factor in the
Be phenomenon is not high rotational velocity itself but its ratio to
some critical velocity, which measures the tendency of the star to be
centrifugally disrupted (e.g., Porter 1996; Keller et al. 1999; 2001).
More complicated evolutionary models can therefore be envisaged in
which loss of angular momentum via wind breaking stops the Be
phenomenon (cf. Meynet \& Maeder 2000; Keller et al. 2001).

In any case, evidence from young open clusters in the Milky Way and
the SMC (e.g.,  
Keller et al. 1999; 2001) supports the idea that fast rotating B stars
become Be stars late in their MS life, as the highest fraction of Be
stars appears to occur at the MS turn-off and above it. Therefore,
whatever the mechanisms involved, if
the Be phenomenon occurs preferentially in stars that are close to the
end of their MS lives or have already started to move away from the MS
and, at the same time, we know that it has already stopped when the
star reaches the supergiant stage, it is
tempting to conclude that the Be phenomenon is related to some
physical conditions occurring only during a fraction of the lifetime
of the star. If this is so,
some information on the mechanism producing the Be phenomenon can be
gained by determining at what phase in the  
evolution of the star the Be characteristics disappear and what are the
physical parameters of the star when this happens.

In this spirit, a programme has been started to identify the highest
luminosity at which a star may display the Be phenomenon.
If one can identify stars of relatively high luminosity 
which still display the Be phenomenon, important constraints could
be gained on the physical conditions that lead to its
development. Some care is necessary here, as Be characteristics are
mainly observational, and the assumption that they obey to a single
physical mechanism may be misleading. Therefore a careful
determination of the stellar parameters of putative high luminosity Be
stars via fitting of stellar atmosphere models will be necessary
before any conclusion can be reached.

As a first step for such work, in this paper I set out to investigate the
existence of high-luminosity Be stars. As Be stars are known to be
widespread among luminosity classes III to V, attention is centred on
OB stars of luminosity class I or II that may display Be characteristics.

\section{Target selection and Observations}

A search of SIMBAD was made in order to select objects classified as
 emission-line stars with spectral types earlier than A0 and
 luminosity classes I or II. Only objects North of $\delta=-10\degr$
 were considered because of the telescopes used. Stars with luminosity
 class Ia were excluded, as many of them appear in lists of
 emission-line stars, even when their H$\alpha$ emission lines are
 morphologically normal for their spectral type; for example, HD~14818
 (= MWC 45) or HD~224055 (= MWC~404) in Fig~\ref{fig:superalpha}.

\begin{figure}[ht]
\resizebox{\hsize}{!}
{\includegraphics[bb=80 180 500 600]{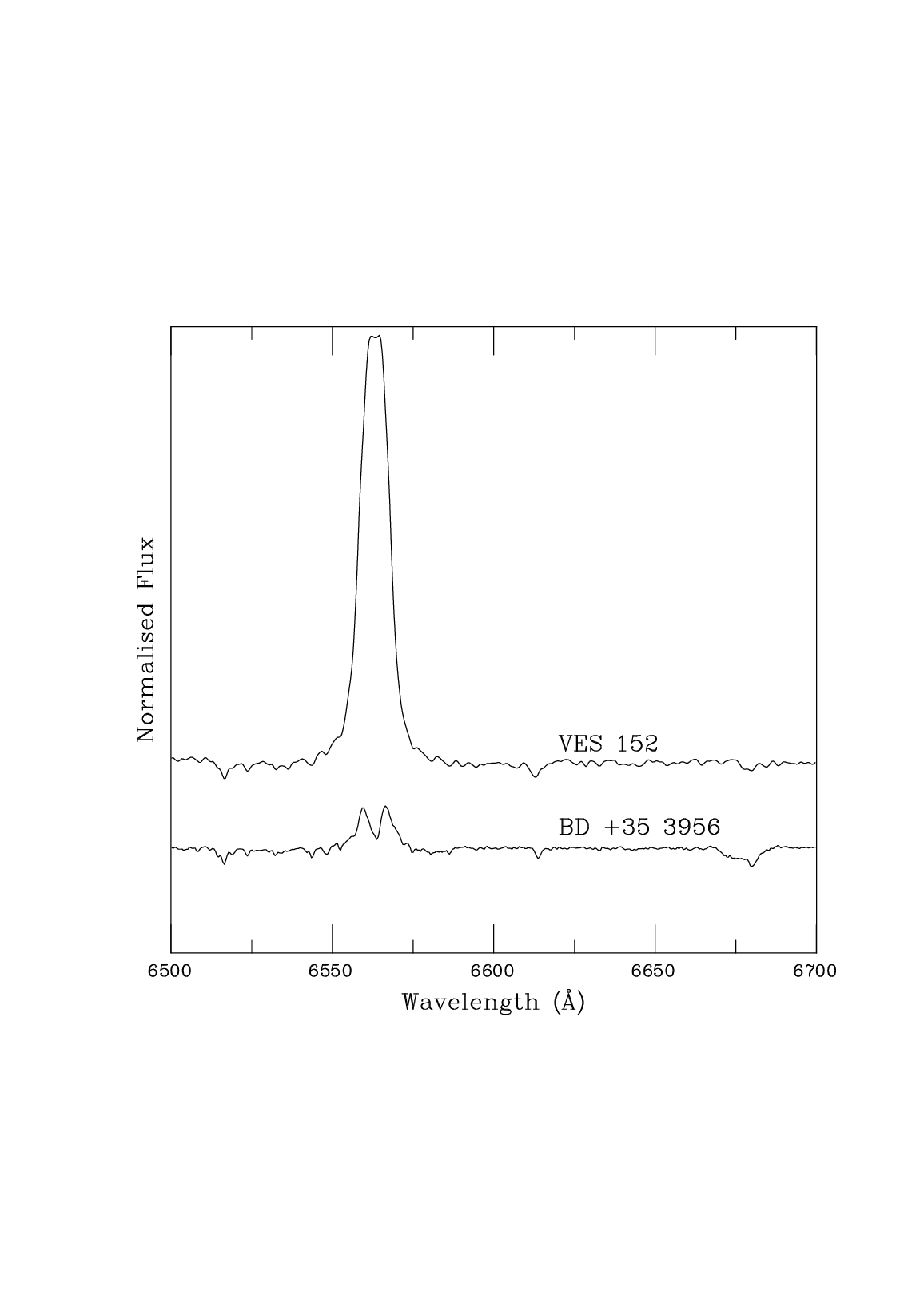}}
\caption{Typical H$\alpha$ emission lines in Be stars. BD
$+35\degr$3956, discussed in Section~\ref{sec:faint}, is a weak Be
star, and the emission line clearly displays the double peaks due to
Keplerian broadening. In Be stars with stronger emission, like VES~152
(another Be star in the field of NGC~6871), the emission line is
optically thick and the double-peaked structure can only be
appreciated at relatively high resolution if the inclination to the
line of sight is moderate or low.}
\label{fig:minialpha}
\end{figure}

A literature search showed that many O-type stars classified as
emission-line objects had later been considered morphologically normal
Of stars. The same applies to several B stars or luminosity classes
Iab and Ib which display P-Cygni emission profiles in H$\alpha$. Also,
it was seen that many stars classified as late-B/A luminosity class II
objects in old studies (such as the LS catalogue; Hardorp et al. 1959,
and continuations) were later found to be shell stars of much lower
luminosity. A shell star is a Be star seen at high inclination, which,
in addition to emission lines, presents sharp absorption cores of
singly-ionised metals, which at low resolution, may resemble a
high-luminosity star (cf.~Porter 1996; Hanuschik 1995).

\begin{table*}[ht]
\caption{Log of observations. Objects specifically targeted because
they have been referred to in the literature as high-luminosity emission-line
stars are marked with a $\dagger$. Other objects are comparison stars
in the same clusters.}
\label{tab:obs}
\begin{tabular}{llccc}\hline
Association&Object &Date & Configuration & Spectral Range\\ 
\hline
Vul OB1 (NGC 6823) &HD 344863$^{\dagger}$&Jul 1st, 2003 & INT+IDS+R900V& 3500--5850\AA\\
&&Jul 3rd, 2003 & INT+IDS+R900V& 4900--7200\AA\\
&HD 344873$^{\dagger}$&Jul 2nd, 2003 & INT+IDS+R900V& 3500--5850\AA\\
&&Jul 3rd, 2003 & INT+IDS+R900V& 4900--7200\AA\\
Cyg OB3 (NGC~6871)&HD 190918 &Jul 31st, 2001 & 1.52+R1200 &3950--4410\AA\\
&&Aug 1st, 2001 & 1.52+R1200 &4460--4910\AA\\
&HD 227611$^{\dagger}$&Jul 18th, 2001& 1.93+R300 & 3800--6850\AA\\
&&Jul 31st, 2001 & 1.52+R1200 &3950--4410\AA\\
&&Aug 1st, 2001 & 1.52+R1200 &4460--4910\AA\\
&&Aug 2nd, 2001 & 1.52+R1200 &6380--6830\AA\\
&HD 227634&Jul 31st, 2001 & 1.52+R1200 &3950--4410\AA\\
&&Aug 1st, 2001 & 1.52+R1200 &4460--4910\AA\\
&HD 227733$^{\dagger}$&Aug 1st, 2001 & 1.52+R1200 &4460--4910\AA\\
&&Jul 6th, 2002 & 1.93+R1200& 3960--4880\AA\\
&BD $+35\degr$3955&Jul 31st, 2001 & 1.52+R1200 &3950--4410\AA\\
&&Aug 1st, 2001 & 1.52+R1200 &4460--4910\AA\\
&BD $+35\degr$3956$^{\dagger}$&Jul 31st, 2001 & 1.52+R1200&3950--4410\AA\\
&&Aug 1st, 2001 & 1.52+R1200 &4460--4910\AA\\
&&Jul 6th, 2002 & 1.93+R1200& 3960--4880\AA\\
&&Oct 1st, 2002 & 1.93+R1200& 6250--7140\AA\\
Cyg OB1 (Be 87)&HD 229059$^{\dagger}$&Jul 5th, 2003 & INT+IDS+R900V& 3500--5850\AA\\
&&Jul 5th, 2003 & INT+IDS+R900V& 4900--7200\AA\\
&BD $+36\degr$4032&Jul 4th, 2003 & INT+IDS+R900V& 3500--5850\AA\\
&&Jul 5th, 2003 & INT+IDS+R900V& 3500--5850\AA\\
&&Jul 6th, 2003 & INT+IDS+R900V& 4900--7200\AA\\
&V439 Cyg$^{\dagger}$ & Oct 23rd, 2001 & 1.93+R600 & 3750--5550\AA\\
&&Oct 22nd, 2001 & 1.93+R1200 & 6250--7140\AA\\
Cyg OB1&HD 194280$^{\dagger}$&Jul 1st, 2003 & INT+IDS+R900V& 3500--5850\AA\\
&&Jul 5th, 2003 & INT+IDS+R900V& 4900--7200\AA\\
&HD 194334$^{\dagger}$&Jul 1st, 2003 & INT+IDS+R900V& 3500--5850\AA\\
Cyg OB1 (NGC 6913) & HD 229221$^{\dagger}$&Jul 4th, 2003 & INT+IDS+R900V& 3500--5850\AA\\
&HD 229227 &Jul 4th, 2003 & INT+IDS+R900V& 3500--5850\AA\\
&HD 229234 &Jul 3rd, 2003 & INT+IDS+R900V& 3500--5850\AA\\
&HD 229238 &Jul 3rd, 2003 & INT+IDS+R900V& 3500--5850\AA\\
&HD 229239$^{\dagger}$ &Jul 3rd, 2003 & INT+IDS+R900V& 3500--5850\AA\\
&&Jul 5th, 2003 & INT+IDS+R900V& 4900--7200\AA\\
None&HD 207329$^{\dagger}$&Jul 7th, 2003 & INT+IDS+R900V& 3500--5850\AA\\
&&Jul 7th, 2003 & INT+IDS+R900V& 4900--7200\AA\\
Cam OB3 & LS I $+56\degr$92$^{\dagger}$ &Jul 23rd, 2002 & INT+IDS+R1200Y&4050--4905\AA\\
&&Jul 25th, 2002 & INT+IDS+R1200Y&6270--7120\AA\\
\hline
\end{tabular}
\end{table*}

It has also been suggested that some low-mass interacting binaries and
 post-AGB stars may display spectra
 very similar to those of supergiants during short periods of time
 (e.g., Saselov 1986). For these reasons, targets were preferentially
 selected 
 among objects belonging to open clusters or associations, so as to
 make sure that they were really massive stars of high
 luminosity. This selection made also possible to observe other
 members of these associations to be used as comparison. 


Though many of the clusters considered are relatively well studied and
spectral classifications exist for some of their members, it was felt
necessary to re-observe some of the brightest members in order to
provide a consistent framework. As an example, Table~\ref{tab:m29}
lists the spectral types assigned by different authors to the five
bright stars defining the core of NGC~6913, which is thought to be
a Trapezium-like system. Even though there is a {\it general}
agreement, there is a lot of variation. Apart from some systematic
effects (the intermediate types B0.2 and O9.7 were not defined at the
time of the first works), this variation is most likely caused by
differences of appreciation by the various authors. As we are concerned
here with small differences in luminosity, having all spectral types
in a single {\it consistent} system will allow us to effectively
compare the Be stars with other luminous members.

\begin{table}[htb]
\caption{Spectral types assigned to the five bright members forming
the core of NGC~6913 by different authors. References are R51: Roman
(1951); M53: Morgan et al. (1953); M95: Massey et al. (1995); WH00:
Wang \& Hu (2000); Here: this work.}
\label{tab:m29}
\begin{tabular}{lccccc}\hline
Object & \multicolumn{5}{c}{Spectral Type}\\ 
 & R51 &M53 & M95 & WH00 & Here \\
\hline
HD 229221 & B0\,II?e & B0?\,I?pe & Be & B0\,IIIe & B0.2\,IIIe\\
HD 229227 & B0\,III & B0\,II & B0.2\,III & B0\,II & O9.7\,III\\
HD 229234 & O9\,II & O9.5\,III & O9\,If & O7\,II & O9\,II\\
HD 229238 & B0.5\,II & B0.5\,Ib & B0\,I & B0\,I & B0.2\,II\\
HD 229239 & B0.5\,IIe &B0.5\,Iab & B0\,IV & B0\,I & B0.2\,III\\
\hline
\end{tabular}
\end{table}

Observations for this project have been carried out with several
telescopes, mostly as poor-weather backup for other projects, taking
advantage of the brightness of targets. Most of the observations were
obtained during a run on 2003 July 1st--7th at the 2.5-m Isaac Newton
Telescope (INT) at La Palma, Spain. 
The telescope was equipped with the Intermediate Dispersion
Spectrograph (IDS) and the 235-mm camera. The R900V grating and
EEV\#13 camera were used, resulting in a nominal dispersion of
$\approx0.65$\AA/pixel. Further INT observations had been obtained on
2002 July 23rd--25th, using a similar configuration, but this time
with the R1200Y grating and Tek\#4 camera, which gives a dispersion of
$\approx0.8$\AA/pixel. 

Observations of stars in NGC~6871 were obtained as backup of several
runs at the Observatoire de Haute Provence (OHP) during 2001 and
2002. The brightest stars were observed with the {\em Aur\'{e}lie}
spectrograph on the 1.52-m telescope during a run on 2001 July
31st--August 2nd. The
spectrograph was equipped with grating \#3 (600 ln mm$^{-1}$) and
the Horizon2000 $2048\times1024$ EEV CCD camera, resulting in a
dispersion of 0.22\AA/pixel (resolving power of approximately 7000).
Other stars were observed with the 1.93-m telescope 
equipped with the long-slit  
spectrograph {\em Carelec} and the EEV CCD. 
 
The complete list of observations, with details of the configurations
used, is shown in Table~\ref{tab:obs}.

Spectral classification in the MK system is based on comparison with
standard stars. The standards used here are those from Steele et
al. (1999), which were observed with the INT + IDS + R1200B + EEV\#12,
resulting in a dispersion of $\sim0.5$\AA/pixel, i.e., slightly higher
than with the R900V grating used for most of our observations. They have
been complemented with spectra of \object{$\iota$ Ori} (O9\,III),
\object{$\upsilon$ Ori} (B0\,V), \object{HD~48434} (B0\,III) and $\iota$
CMa (B3\,II) taken with the ESO 1.52-m telescope at La Silla
Observatory, Chile, equipped with  
the Boller \& Chivens spectrograph, fitted with
the \#32 holographic grating and the Loral \#38 camera. This
configuration also 
gives a nominal resolution of $\sim 0.5 $\AA/pixel

The set of standards has been complemented by interpolation based on
the grid provided by Walborn \& Fitzpatrick (1990). Within the
spectral range considered, the spectral type is mainly determined by
the ratios of \ion{Si}{iii} and \ion{Si}{iv} lines, while luminosities
are deduced from the ratios between these lines and the \ion{He}{i}
lines (Walborn 1971). 

\section{Results}

\subsection{NGC~6823}

This relatively young cluster in the Vul OB1 association contains the 
O7\,V((f)) star HD~344784 and the B0.5\,Ib supergiant HD~344776. Two cluster
outliers, believed to belong to the association, have been suggested as
high-luminosity Be stars.

\begin{figure*}[ht]
\resizebox{\hsize}{!}
{\includegraphics[bb= 50 130 350 710,angle=-90]{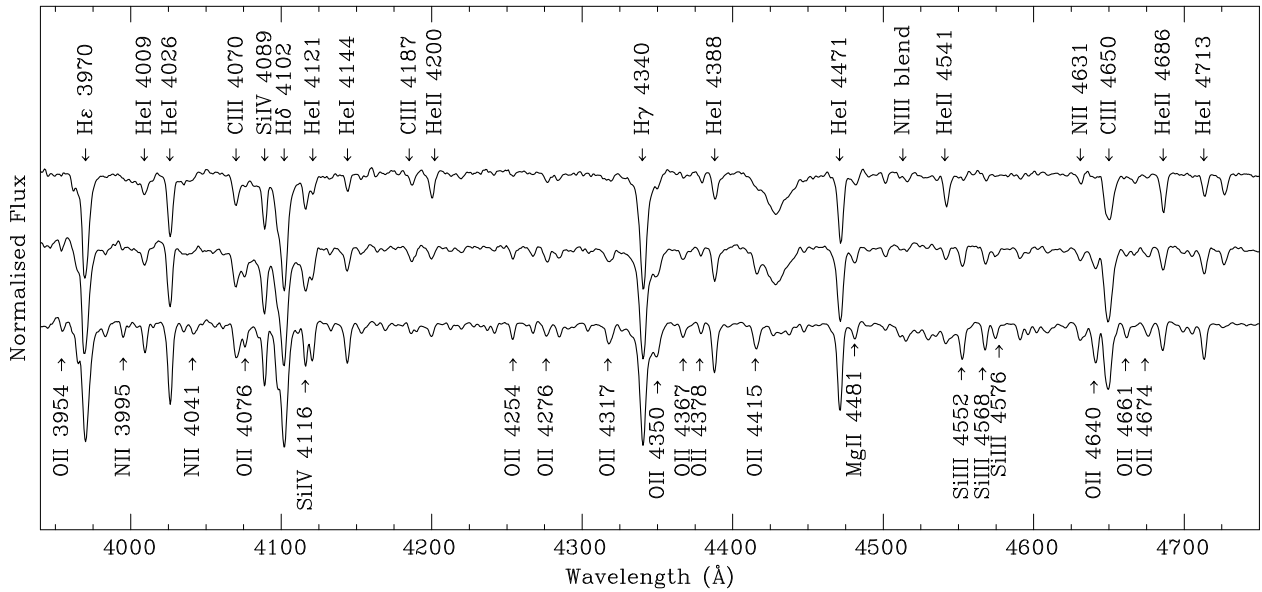}}
\caption{Classification spectra of two outliers of NGC~6823, reported
in the literature to have shown a Be phase, HD~344863 (top) and
HD~344873 (middle). The bottom spectrum is that of the B0\,III MK
standard \object{HD 48434}, artificially spun up to $120\,{\rm
km}\,{\rm s}^{-1}$, for comparison.}
\label{fig:outblue}
\end{figure*}

\subsubsection{HD 344873}
This star was originally classified as B0\,II by Morgan et al. (1953), though 
its luminosity was considered uncertain by Walborn (1971), who gave it
as B0\,III?((n)). Massey et al. (1995) classified it as B0\,III.
HD~344873 is listed in Wackerling's (1970) catalogue of emission-line
stars, as it appears as OB$^{+}$h,r (H$\alpha$ in emission) in the LS
catalogue (LS II $+23\degr$43).  

In Fig.~\ref{fig:outblue}, the spectrum of HD~344873 appears compared
to that of the B0\,III 
standard HD~48434. The two spectra are very similar, though the
\ion{Si}{iii} and specially the \ion{O}{ii} lines in HD~344873 are
clearly weaker. This is suggestive of an earlier spectral type and,
making use of the condition
\ion{He}{ii}~$\lambda$4541\AA$\simeq$\ion{Si}{iii}~$\lambda$4552\AA,
we can consider a classification O9.7 for HD~344873. The weaker
\ion{He}{ii}~$\lambda$4686\AA\ would then suggest a higher luminosity,
and this is supported by the strong \ion{Si}{iv} doublet. Therefore we
take O9.7\,II as the spectral type for HD~344873.

The H$\alpha$ region of HD~344873 is shown in
Fig.~\ref{fig:notseen}. There is no hint of emission components.

\subsubsection{HD 344863}
Originally classified as B0\,II? by Morgan et al. (1953), HD~344863
has been marked as a Be star by several authors. It appears as
OB$^{+}$ce,h,r (H$\alpha$ and Balmer discontinuity in emission) in the
LS catalogue (LS II $+24\degr$12) and is given as B0?\,II-IIIe by 
Schmidt-Kaler (1967), but is reported to be strongly in absorption in
1982 (EW$_{\rm{H}\alpha}=+3.3$\AA) by Peppel (1984). 

The spectrum of HD~344863 is clearly {\em earlier} than that of
HD~344873. \ion{O}{ii} lines are basically absent from the spectrum,
as also is the \ion{Si}{iii} triplet. \ion{He}{ii}~$\lambda$4541\AA\
is well developed, but much weaker than \ion{He}{i}~$\lambda$4471\AA,
suggesting a spectral type around O9. The fact that
\ion{He}{ii}~$\lambda$4541\AA $>$ \ion{He}{i}~$\lambda$4387\AA\, and
\ion{He}{ii}~$\lambda$4200\AA $>$ \ion{He}{i}~$\lambda$4144\AA\ would
allow an O8.5 classification (N. Walborn, priv. comm.). As
\ion{He}{ii}~$\lambda$4686\AA\ is 
not stronger than \ion{C}{iii}~$\lambda$4650\AA, it must have a
relatively high luminosity. So we adopt O9\,III for HD~344863, though
it is close to O8.5.

\begin{figure}[htb]
\resizebox{\hsize}{!}
{\includegraphics[bb= 60 160 530 620]{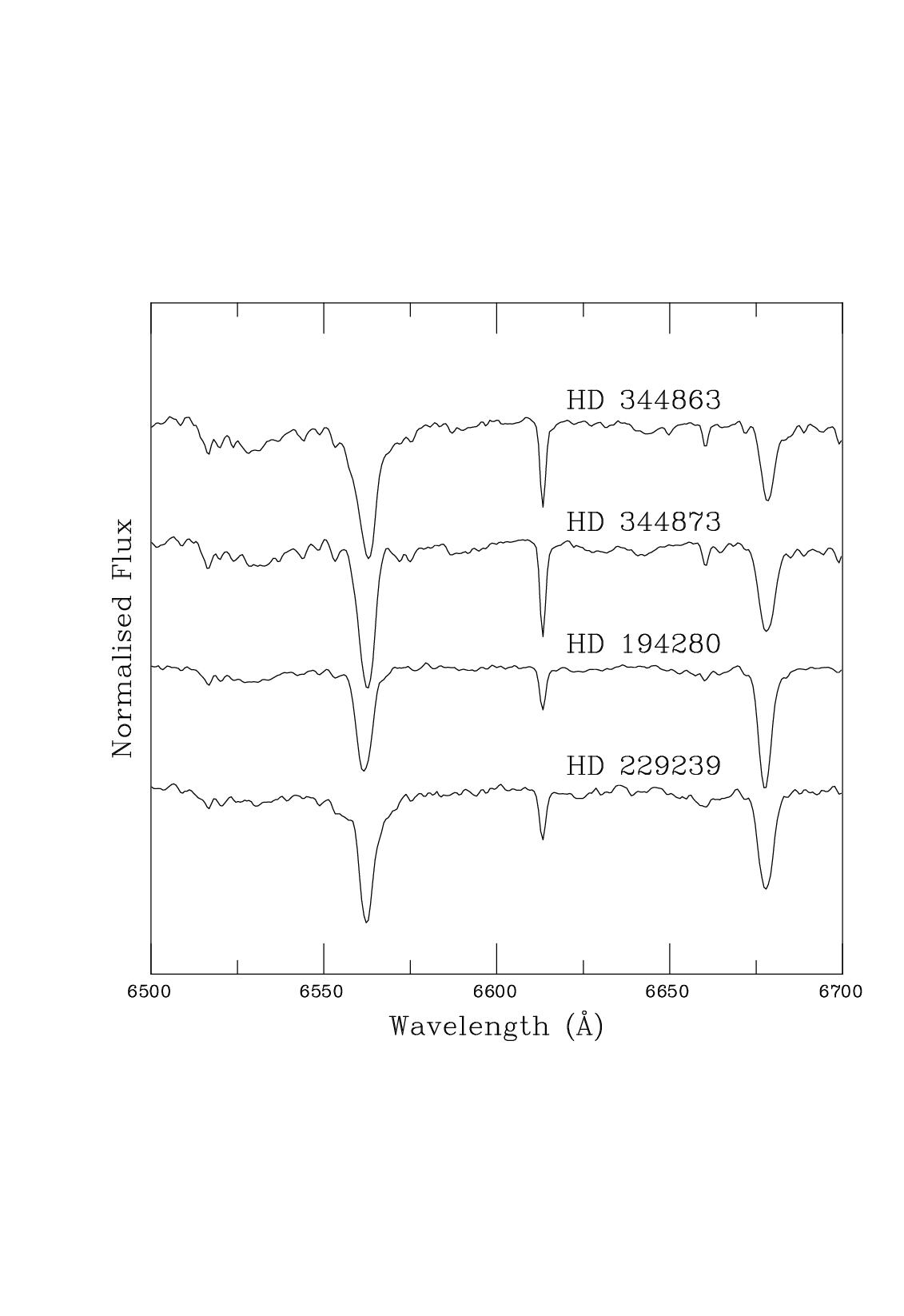}}
\caption{The H$\alpha$ region for four luminous stars reported in the
literature to have shown emission lines, but appearing normal now.}
\label{fig:notseen}
\end{figure}

The shape of H$\alpha$ (Fig.~\ref{fig:notseen}) shows that HD~344863
is not in a Be phase at present.

\subsection{NGC 6871}
NGC~6871 is a rather extended open cluster, whose boundaries are not
well defined. It is believed to be the core of the Cyg OB3
association. Estimates of its distance and age vary by relatively
large amounts between different authors. Recent searches of
emission-line stars have been carried out by Bernabei \& Polcaro
(2001) and Balog \& Kenyon (2002), resulting in the detection of 
several faint Be stars and some pre-main-sequence objects.

\begin{figure*}[ht]
\resizebox{\hsize}{!}
{\includegraphics[bb= 100 120 440 700,angle=-90]{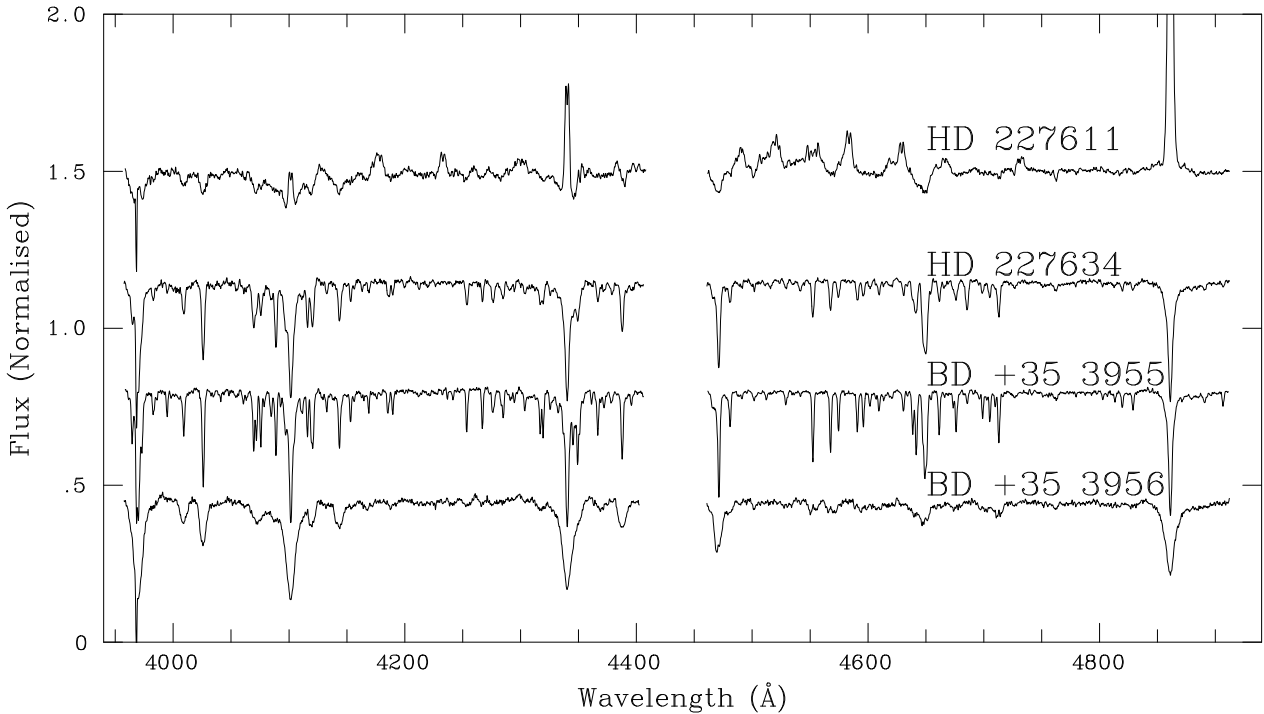}}
\caption{The two brightest Be stars in NGC~6871 compared to two other
bright stars at the cluster centre. Apart from the obvious difference in the
rotational velocities, the Be stars appear clearly to be of rather
lower luminosity. The four spectra displayed here have been obtained
with the 1.52-m telescope at OHP and have resolutions more than twice
better than the spectra in Figures~\ref{fig:outblue}, \ref{fig:core}
and~\ref{fig:sgs}. The small gap around
   $\lambda = 4420$\AA\ indicates the division between two
   poses.} 
\label{fig:dwarves}
\end{figure*}

The brightest star in the cluster is the O9.5\,I+WN4 binary
HD~190918. The spectra of four other bright stars in the cluster core 
are displayed in Fig.~\ref{fig:dwarves} (though HD~227611 is actually
a few arcmin apart). 

\subsubsection{BD +35$\degr$3956}
\label{sec:faint}

BD $+35\degr$3956, classified B0.5\,V by Morgan et al. (1953), was found
to be a mild Be star by Grigsby \& Morrison (1988). Its H$\alpha$
spectrum, shown in Fig~\ref{fig:minialpha}, confirms that it is still
in the Be phase. The apparent brightness of BD $+35\degr$3956 compared
to nearby stars known to be of high luminosity has 
prompted its inclusion in this work.

The classification spectrum of BD $+35\degr$3956 is shown in
Fig.~\ref{fig:dwarves}. All the lines are very broad. This is likely
to be partly due to fast rotation, but
the shape of the metallic lines strongly suggests that it is a
double-lined spectroscopic binary, though it cannot be ascertained
with this S/N and resolution. Considering the weakness of
\ion{He}{ii}~$\lambda4686$\AA, the star has to be B0.5 or even
later. As \ion{Si}{iv}~$\lambda4089$\AA\ is hardly seen, it is a
main-sequence star. So the classification by Morgan et al. (1953) is
confirmed and this object is not a luminous Be star.

\subsubsection{HD~227611}
First quoted as an emission-line star by Merrill et al. (1925),
HD~227611 (= MWC 327) was classified Bpe by Roman (1951) and B0\,II?pe
by Morgan et al. (1953), the
peculiarity likely referring to the strength of its emission lines. 

HD~227611 is a very extreme Be star. In its red spectrum, I measure
EW$_{\rm{H}\alpha}=-88$\AA, which is close to the upper limit of
values reported for Be stars. Its blue spectrum, shown in
Fig.~\ref{fig:dwarves} is covered with 
double-peaked emission lines, corresponding to \ion{Fe}{ii} and other
metals. Its spectral type and luminosity are therefore very difficult
to determine. Considering that no \ion{He}{ii} lines are visible, it
must be later than B0. As none of the \ion{Si}{iv} doublet lines
is comparable to the neighbouring \ion{He}{i} lines (they are not even
seen), but the
\ion{O}{ii}/\ion{C}{iii} complex at $\lambda\lambda6450-6550$ is
prominent, it must be a 
main-sequence star earlier than B1 or a giant close to B1. It is
certainly not a high-luminosity Be star.

\subsubsection{HD~227733}

A slightly fainter star in NGC~6871, HD~227733 was found to be an
emission-line star during the Vatican 
survey (MacConnell \& Coyne 1983). Bernabei \& Polcaro (2001)
suggested that it was surrounded by a bright nebula, prompting its
study here. 

The spectrum of
HD~227733 is unremarkable, and it is not shown. Its spectral type
is B1.5\,Ve. No emission was present in H$\beta$ in the 2001
observation, but weak emission peaks appear inside the line in the
2002 spectrum.

\subsubsection{HD~227634 and BD +35$\degr$3955}

Two of the brightest stars in NGC~6871 have been observed
for comparison. Their spectra are displayed in Fig.~\ref{fig:dwarves}.
Other stars of similar apparent brightness are HD~190864 --
O6.5\,III(f) -- and HD~190919  -- B0.7\,Ib  (Walborn 1971).

HD 227634 displays moderately strong \ion{He}{ii}~$\lambda4686$\AA,
which makes it earlier than B0.5 and
\ion{Si}{iv}~$\lambda4089$\AA$\gg$\ion{He}{i}~$\lambda4121$\AA, which
makes it a luminous star. Comparison with other objects indicates that
its spectral type is B0.2\,II, though one would think it is slightly
earlier and less luminous than HD~229239, discussed later. Walborn
(1971) does indeed classify it B0\,II.

BD $+35\degr$3955 displays only a barely visible
\ion{He}{ii}~$\lambda4686$\AA\ line, which puts it at B0.7. The
narrowness of its lines and lack of Stark broadening in the Balmer
lines indicates that it is very luminous. The strength of the
\ion{O}{ii} and \ion{Si}{iii} lines puts it at B0.7\,Iab, close to
Ia, in excellent agreement with Walborn (1971).

\subsection{NGC 6913}

This young open cluster in the Cyg OB1 association, also known as M~29,
contains a large number of luminous stars with 
spectral types around B0, several of which have been reported to present
emission lines. As such, it represent a unique opportunity to
correctly estimate the luminosities of the brightest Be stars.

\subsubsection{HD 229221}
Given as an emission line star (MWC 344) by Merrill \& Burwell (1933), 
HD~229221 was classified B0\,II?e by Roman (1951) and B0?\,I?pe by Morgan et
al. (1953). HD~229221 appears to have displayed emission lines at all
epochs, as it is given as B0e by Walker \& Hodge (1968), mentioned to
display EW$_{\rm{H}\alpha}=-13$\AA\ by Schild \& Romanishin (1976), and 
classified Be by Massey et al. (1995). Wang \& Hu (2000) classify it
as B0\,IIIe.

\begin{figure}[ht]
\resizebox{\hsize}{!}
{\includegraphics[bb= 50 140 320 710,angle=-90]{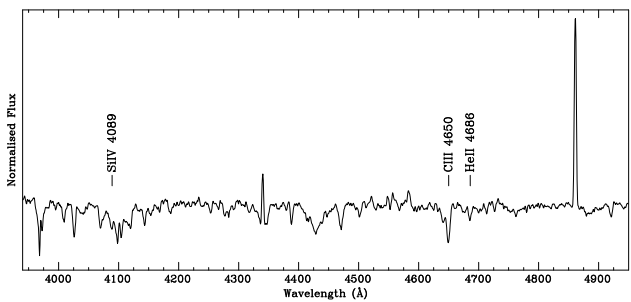}}
\caption{The blue spectrum of the Be star HD~229221, in NGC~6913. The
strength of the emission lines is typical of a Be star. The star is
moderately luminous, as metallic lines are clearly seen in spite of
the relatively fast rotation.}
\label{fig:bigbe}
\end{figure}

The classification of HD~229221 is complicated by the relatively
strong emission spectrum. The weak \ion{He}{ii}~$\lambda$4686\AA\ and
absence of \ion{He}{ii}~$\lambda$4200\AA\ make it later than B0. On
the other hand, \ion{C}{iii}~$\lambda$4650\AA\ is very strong when
compared to \ion{O}{ii} lines, suggesting a spectral type B0.2. The
weakness of \ion{He}{ii}~$\lambda$4686\AA\ indicates then a relatively
high luminosity, but \ion{Si}{iv}~$\lambda$4089\AA\ does not appear
very strong compared to the \ion{He}{i} lines, preventing it from
being higher than III. We take then B0.2\,IIIe.

\subsubsection{HD 229239}
HD~229239 is given as an emission line star (MWC 1016) by Merrill \& 
Burwell (1949), based on an observation obtained in 1946.
It was classified B0.5\,IIe by Roman (1951), but later is given as
B0.5\,Iab by Morgan et al. (1953), without reference to
emission. Morgan et al. (1955) give it as B0\,II and the luminous star
catalogue lists it as OB,r (LS II $+38\degr$82), again without any
reference to emission.

Further reference to HD~229239 as an emission line star is given by
Perraud \& Pelletier (1959), but no reference to emission is made by 
Walker \& Hodge (1968). Unfortunately, none of these sources provides dates
for the observations. No mention of emission lines is found in modern
works. Massey et al. (1995) classify HD~229239 as B0\,IV, while Wang \& Hu
(2000) give B0\,I.

The H$\alpha$ spectrum of HD~229239 is shown in
Fig.~\ref{fig:notseen} and clearly demonstrates that the star is not
in a Be phase at present.

The classification spectrum of HD~229239 is shown in Fig.~\ref{fig:core}.
The weak \ion{He}{ii}~$\lambda$4686\AA\ and absence of
\ion{He}{ii}~$\lambda$4200\AA\ make it later than B0, the ratio of
\ion{Si}{iii} to \ion{Si}{iv} suggesting B0.2. The strength of the
\ion{Si}{iv} compared to the neighbouring \ion{He}{i} lines indicates
then a luminosity class III, in agreement with the strength of
\ion{He}{ii}~$\lambda$4686\AA. Therefore we adopt B0.2\,III.

\subsubsection{HD 229227, HD 229234 and HD 229238}
\begin{figure*}[ht]
\resizebox{\hsize}{!}
{\includegraphics[bb= 50 140 320 710,angle=-90]{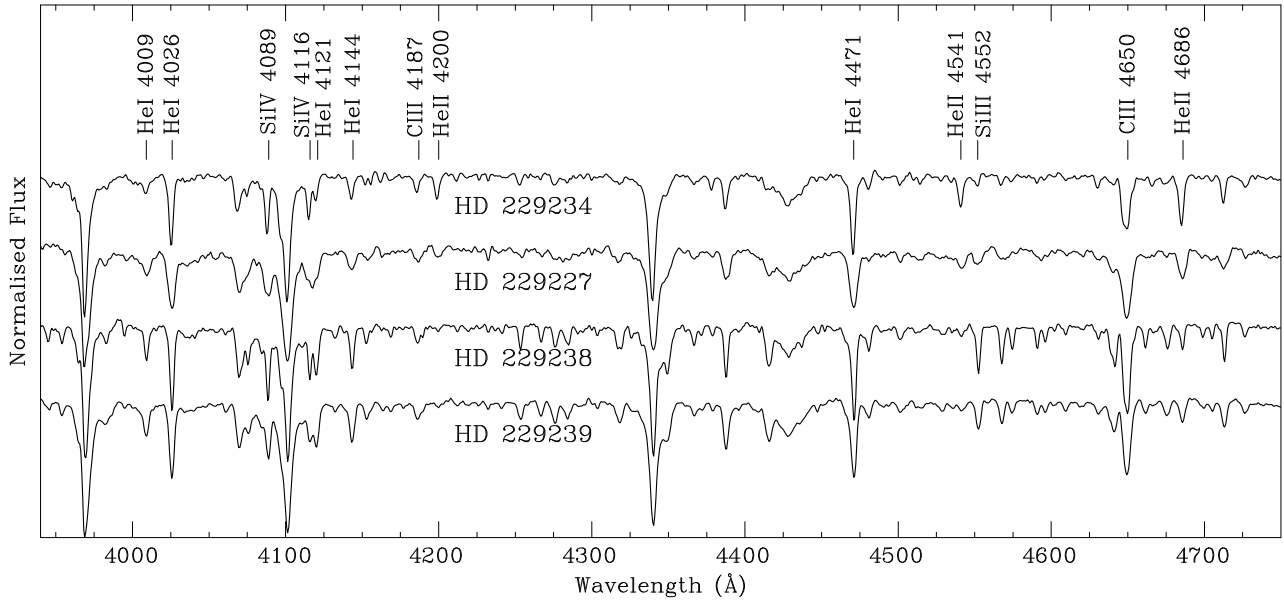}}
\caption{Classification spectra of four bright stars at the core of
NGC~6913. HD~229227, which is clearly a very fast rotator, has never
been reported to be a Be star. HD~229239, on the other hand, was in a
Be phase before 1950.}
\label{fig:core}
\end{figure*}

None of these stars has been reported as an emission-line object and
they have been observed as comparison. Together with HD~229221 and
HD~229239, they are the brightest members of NGC~6913 and constitute
its obvious core. Their spectra are displayed in Fig.~\ref{fig:core}.

The lines in HD~229227 are rather broad, suggesting that it is a fast
rotator. The condition 
\ion{He}{ii}~$\lambda$4541\AA$\simeq$\ion{Si}{iii}~$\lambda$4552\AA\
defines spectral type O9.7. The strength of
\ion{He}{ii}~$\lambda$4686\AA\ is rather small, suggesting a
relatively high luminosity. We adopt O9.7\,III.

The spectrum of HD~229234 is very similar to that of HD~344863. The
ratio of \ion{He}{ii}~$\lambda$4541\AA\ to
\ion{He}{ii}~$\lambda$4471\AA\ supports an O9 spectral type. Metallic
lines are moderately stronger than in HD~344863, indicating a slightly
higher luminosity. We adopt O9\,II.

The spectrum of HD~229238 is very similar to that of HD~229239. Again,
the absence of \ion{He}{ii}~$\lambda$4200\AA\ and ratio of Si lines
suggest B0.2. The \ion{Si}{iv} lines, however, are clearly stronger
in comparison to the \ion{He}{i} lines, meaning that HD~229238 is more
luminous. We adopt B0.2\,II.

\subsubsection{HD 194280}
HD~194280 is a member of Cyg OB1, lying at some distance from
NGC~6913. Classified as B0\,Ib by Morgan et al. (1955), HD~194280 was
recognised by Walborn (1972) as a C-rich star and reclassified
OC9.7\,Iab. Since 
then, it has been repeatedly observed by several authors as a reference star 
without any mention of emission characteristics (recent high-quality
spectra are provided by Walborn \& Howarth 2000). In spite of this, it
appears in Wackerling's (1970) catalogue, as it is given as OB,ce,h,r
in the LS catalogue (LS II $+38\degr$73).

Our spectrum of HD~194280 shows no sign of emission in H$\alpha$ (see
Fig.~\ref{fig:notseen}). The spectrum shown by Walborn \& Howarth
(2000) could show a very weak incipient P-Cygni profile, but it would
not have been detected in a low resolution spectrum. The possibility
that HD~194280 has been an emission line star remains dependent on a
single reference.

\subsubsection{HD 194334}
Less than $4\arcmin$ from HD~194280, HD~194334 is another member of
Cyg OB1. It was classified O7.5\,V by Morgan et al. (1955). It is
given as OB,ce,le,r (several emission lines) in the LS catalogue (LS II
$+38\degr$74), and hence as an emission  
line star by Wackerling (1970).

The spectrum of HD~194334 displays strong selective \ion{N}{iii}
emission features, typical of an Of star. The \ion{S}{iv}
lines at $\lambda\lambda 4486, 4504$\AA,
typical of luminous O-type stars are also
present. \ion{He}{ii}~$\lambda$4686\AA\ is weakly in absorption and
\ion{He}{i}~$\lambda$4471\AA$\simeq$ \ion{He}{ii}~$\lambda$4541\AA. The
spectral type is hence O7\,II(f).

Though the ``le'' classification in the LS catalogue generally implies
Balmer emission lines, it seems likely that HD~194334 was classified
as an emission-line star because of the lines typically seen in
O-type stars of its luminosity. As the spectral type of HD~194334 falls
well outside the range traditionally occupied by Be stars, and it is a
high-luminosity O star, I will not take it as a serious Be candidate.

\subsection{Berkeley 87}

This moderately reddened cluster has been little studied, as stars are
rather faint in the blue, because of high extinction. Its parameters
are poorly known, but it is believed to be part of Cyg OB1.

\subsubsection{BD +36$\degr$4032}

This O-type star, lying at the core of Be 87 is given as an
emission-line star in SIMBAD, though no references appear to mention
emission features in this object. Even at this relatively low
resolution, the star appears to be a spectroscopic binary. The lines
are very broad and asymmetric, and their shapes change between the two
spectra available. The composite spectrum has an approximate spectral
type O8.5\,V.

\subsubsection{V439~Cyg}
This emission-line star close to the centre of Be 87 has been
suggested to display strong variability and have had an M-supergiant
spectrum in the past (Polcaro \& Norci 1998). It displays relatively
strong Be features, with 
strong Balmer emission (EW$_{\rm{H}\alpha}=-41$\AA). Unfortunately,
our spectrum is not of very high quality and the presence of metallic
emission lines make exact classification  difficult, but all features
seen are compatible with a spectral type B1.5\,Ve. It is certainly not a
luminous Be star.

\subsubsection{HD 229059}
The brightest member of Be 87, though lying slightly off the core
region, HD~229059 was identified as an emission line star (= MWC 1014)
by Merril \& Burwell (1949). It has been classified B1.5\,Ibe by Roman
(1951) and B1.5\,Iap by Morgan et al. (1953).

\begin{figure*}[ht]
\resizebox{\hsize}{!}
{\includegraphics[bb= 50 140 320 710,angle=-90]{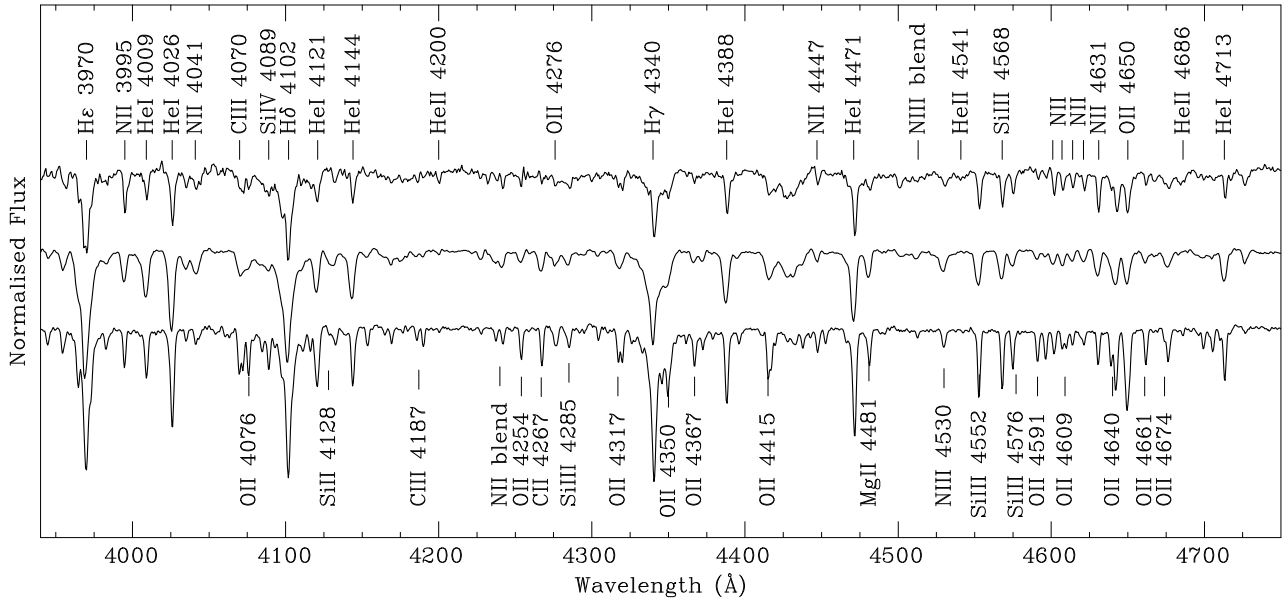}}
\caption{Classification spectra of the two emission-line stars with
apparently higher luminosities found in this survey. The spectrum of
HD~229059 (top) is certainly peculiar, with lines corresponding to a
B0 star, but the \ion{N}{ii} spectrum of a B2 supergiant. HD~207329
(middle) is
a very fast rotator. The spectrum of the B1\,Ib standard HD~24398 has
been artificially spun up to a rotational velocity of
200$\:$km\,s$^{-1}$ (bottom) for comparison.}
\label{fig:sgs}
\end{figure*}

Its spectrum, seen in Fig~\ref{fig:sgs} certainly merits the
``peculiar'' classification. The presence of
\ion{Si}{iv}~$\lambda4089$\AA\ and \ion{He}{ii}~$\lambda$4686\AA\ (the
feature at $\lambda4200$\AA\ could be \ion{He}{ii}  or \ion{N}{ii})
would make it earlier than B1. Its \ion{N}{ii} spectrum, however, is
typical of a B2 supergiant. If one assumes the early spectral type,
the object could be a B0.7\,III star with strong N-enhancement and
moderate C-deficiency. However, the weakness of {\it all} features in
the spectrum of HD~229059, in spite of their sharpness, strongly
suggests the possibility that the spectrum is actually a composite of
a B2\,I supergiant and a lower-luminosity late-O or B0 star. This
possibility would explain the brightness of HD~229059 when compared to
other cluster members (it is almost two magnitudes brighter than BD
$+36\degr$4032). Further support for an O-type star in
HD~229059 is given by the likely presence of
\ion{N}{iii}~$\lambda4097$\AA\ in the wing of H$\delta$.

If we accept the possibility of a BN0.7\,III star, its intrinsic
magnitude would be far too low for it to be a member, when its
reddening clearly supports membership (Turner \& Forbes 1982). 

\begin{figure}[ht]
\resizebox{\hsize}{!}
{\includegraphics[bb= 40 65 560 785]{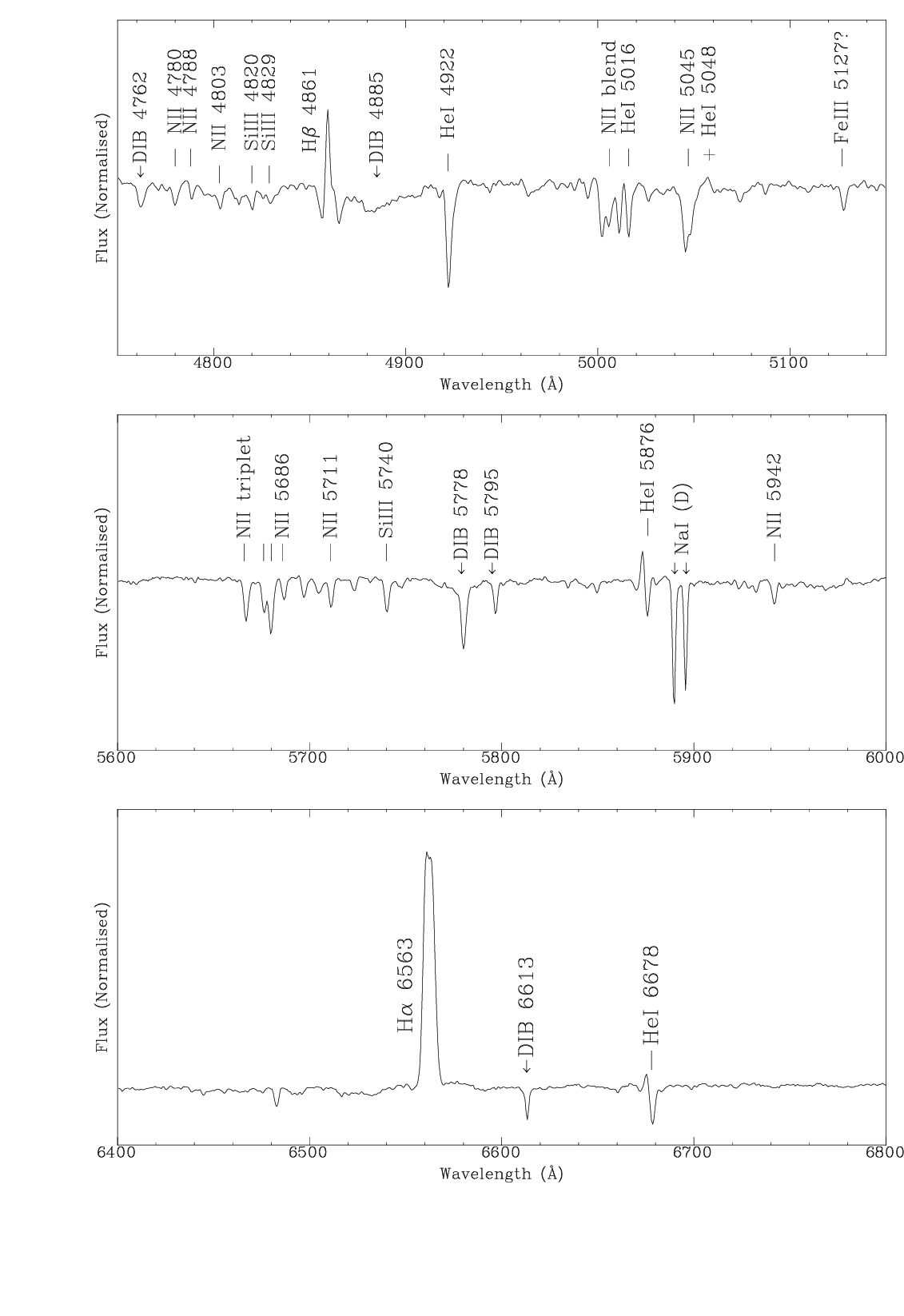}}
\caption{In this montage of different sections of the spectrum of
HD~229059, four emission lines are clearly visible, H$\beta$ (top panel),
\ion{He}{i}~$\lambda5875$\AA\ (mid panel) H$\alpha$ and
\ion{He}{i}~$\lambda6678$\AA\ (both in lower
panel). \ion{He}{i}~$\lambda4922$\AA\ and
\ion{He}{i}~$\lambda5016$\AA\ could show some emission components.  All present the
same symmetry, with the blue peak appearing much stronger. Such
shape could not arise from an expanding envelope.}
\label{fig:super1_lines}
\end{figure}

HD~229059 displays moderately strong emission in H$\alpha$, H$\beta$
and several \ion{He}{i} lines (see Fig.~\ref{fig:super1_lines}). These
lines are clearly asymmetric. H$\alpha$ and H$\beta$ display two
peaks, with a much stronger blue one.

If HD~229059 is a single star, the shape of the lines would indicate
that it is a Be star undergoing V/R variability, as it cannot be
produced by an expanding envelope. A second possibility is that the
lines are produced in a (mildly) interacting binary. In both cases,
the asymmetry of the lines is expected to be variable. Clearly
further observations of this object at higher resolution are needed in
order to settle its nature, the origin of the emission lines and their
possible variability.

\subsection{HD~207329}
Though not an obvious member of any open cluster or association,
HD~207329 seems to be a {\it bona fide} high-luminosity Be
star. Identified as an emission line star (= MWC 378) by Merrill \&
Burwell (1933), it was classified B1.5\,Ib?e by Morgan et al. (1955). It
appears in several works on Be stars, but in the LS catalogue, it is
classified simply as OB (LS III $+51\degr$28) and Walborn (1971)
classifies it as B2.5\,Ia((n)), without reference to emission lines.

\begin{figure}[ht]
\resizebox{\hsize}{!}
{\includegraphics[bb= 40 65 560 785]{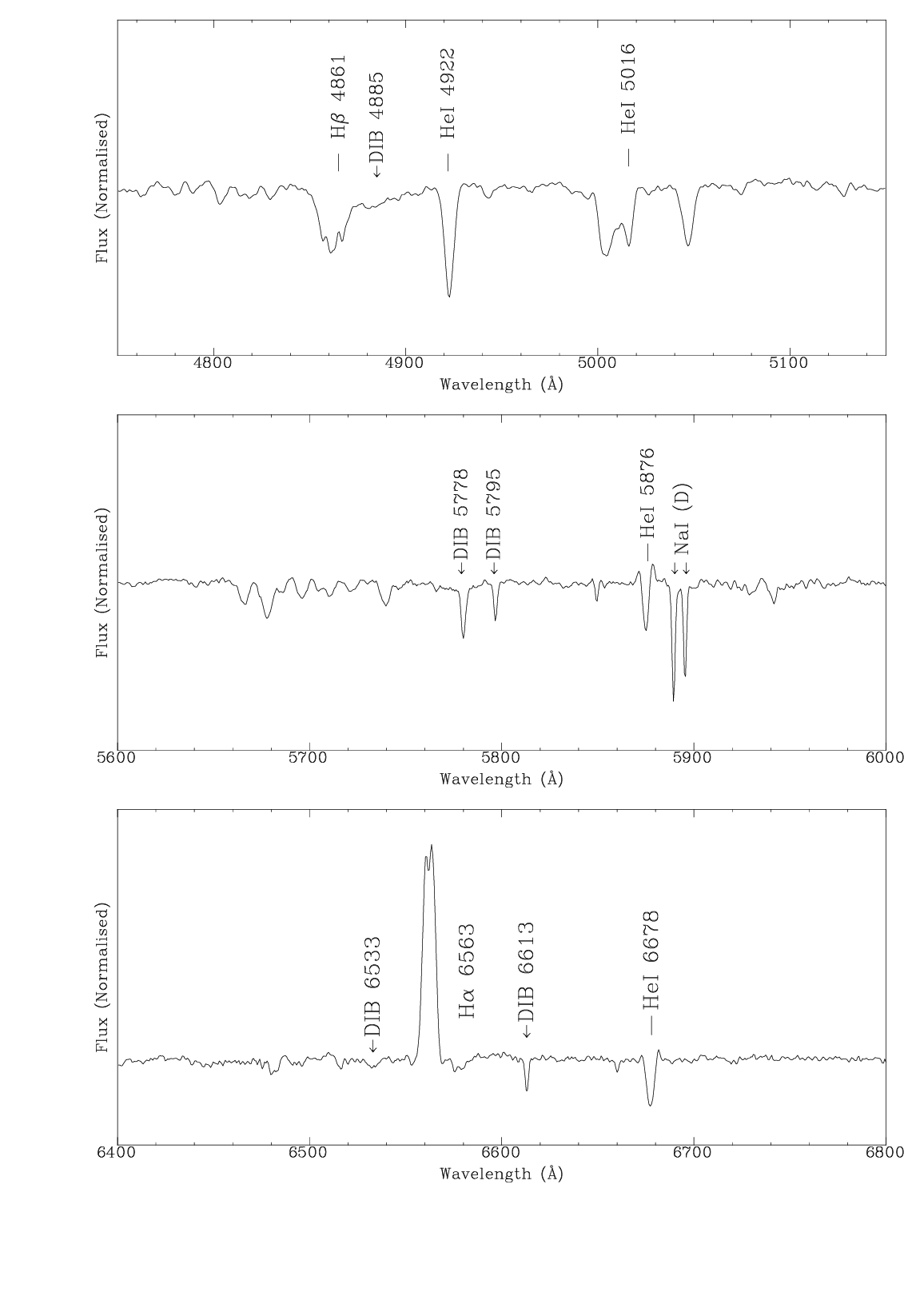}}
\caption{In this montage of different sections of the spectrum of
HD~207329, four emission lines are visible, H$\beta$ (top panel),
\ion{He}{i}~$\lambda5875$\AA\ (mid panel), H$\alpha$ and
\ion{He}{i}~$\lambda6678$\AA\ (both in lower panel). All present the
same symmetry, with two peaks, the red one slightly stronger. Such
shape could arise from either a Keplerian disk or an expanding
envelope. For line identification, see Fig.~\ref{fig:super1_lines}, as
lines in HD~229059 are rather sharper.}
\label{fig:super2_lines}
\end{figure}

As can be seen in Fig.~\ref{fig:sgs} and~\ref{fig:super2_lines},
emission lines in the spectrum of HD~207329 are, with the exception of
H$\alpha$, weak and they may have been missed in low-resolution
spectra, specially if they only covered the blue region.

The classification spectrum of HD~207329, displayed in
Fig~\ref{fig:sgs}, clearly shows that it is a very fast rotator of
high luminosity. Comparison to the B1\,Ib supergiant $\zeta$ Per (which
is artificially spun up to 200$\:$km\,s$^{-1}$ in Fig~\ref{fig:sgs})
indicates that it is of later spectral type, as the
\ion{C}{ii}~$\lambda4267$\AA\ line is clearly stronger
than neighbouring \ion{O}{ii} lines. The fact that
\ion{N}{ii}~$\lambda4631$\AA\ is rather weaker than
\ion{O}{ii}~$\lambda4640$\AA\ makes it earlier than B2 and prevents it
from being N-enhanced. Therefore we take B1.5 as the spectral type.
In spite of the broad profiles, the strength of
\ion{Mg}{ii}~$\lambda4481$\AA\ and the \ion{Si}{iii} lines indicates a
supergiant. I take B1.5\,Ib.

HD~207329 displays moderately strong asymmetric emission lines in
H$\alpha$, H$\beta$ (though the emission does not reach the continuum level)
and several \ion{He}{i} lines (see Fig.~\ref{fig:super2_lines}). The
situation is exactly the opposite as in HD~229059, with the red peak
being slightly dominant. Again, this configuration could be interpreted
as typical of a Be star displaying V/R variations. This kind of
emission lines could also arise in an expanding envelope. This
possibility is directly testable with further observations: if the
star is a Be star (in the sense that its mass loss occurs in the form
of a Keplerian disk), the asymmetry of the lines will change over
time. If the mass loss occurs in a spherically symmetric
configuration, it should not change.

In any case, the fact that HD~207329 is a very fast rotator for its
luminosity seems to support the idea that it is a Be star. If it was
so, it would be the most luminous one known, clearly justifying
further observations.

\subsection{LS I +56$\degr$92}

LS I +56$\degr$92 was observed during a survey for possible members of
the Cam~OB3 association (cf.~Negueruela \& Marco 2003) and found to
display strong emission lines over the spectrum of a luminous mid-B
star.

\begin{figure}[ht]
\resizebox{\hsize}{!}
{\includegraphics[bb= 50 140 320 710,angle=-90]{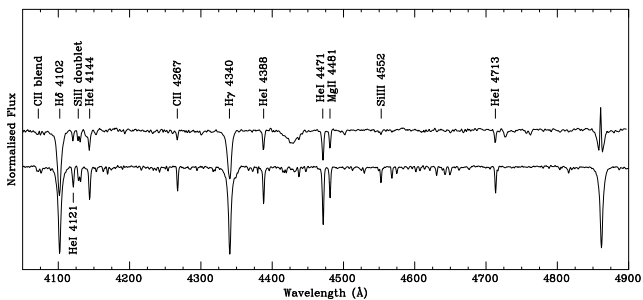}}
\caption{Classification spectrum of LS I +56$\degr$92 (top), showing
the presence of strong single-peaked emission lines. The spectrum of
the B3\,II standard at similar resolution is shown for
comparison. Though LS~I~+56$\degr$92 is obviously later, the Stark
broadening of the Balmer lines is comparable or smaller, indicating a
high-luminosity.} 
\label{fig:superbe}
\end{figure}

 Previous references to this star are scarce. It is listed B6\,Iah in
 the LS catalogue, and hence given as an emission-line star. The only
 other reference quoting spectroscopic observations is McCuskey
 (1956), who gives it as O8 and comments ``very weak H lines''. The
 discrepancy between the two spectral types is certainly surprising,
 as LS I +56$\degr$92 lies in a region of the sky where confusion is
 extremely unlikely.

\begin{figure}[ht]
\resizebox{\hsize}{!}
{\includegraphics[bb=60 165 520 600]{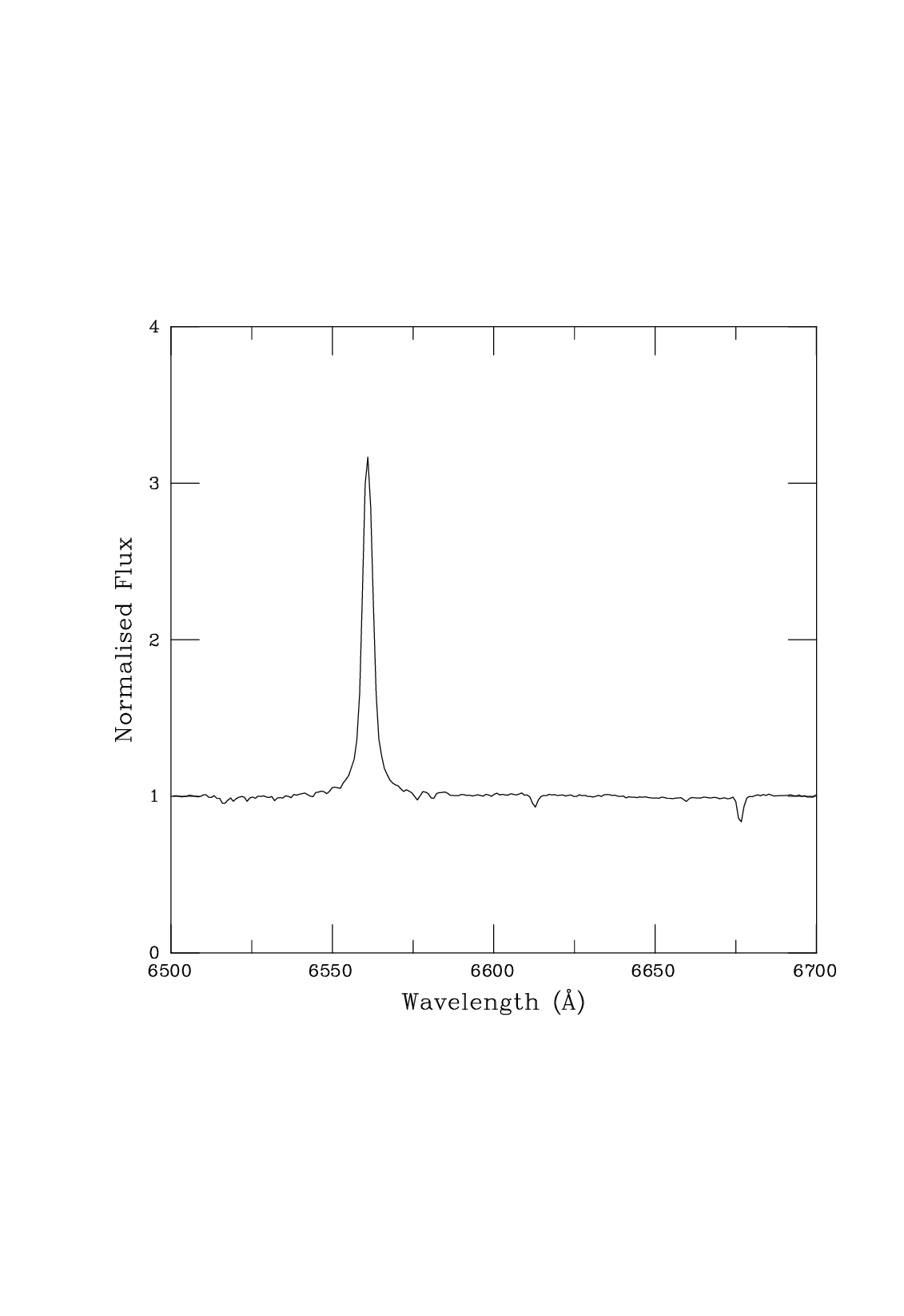}}
\caption{H$\alpha$ region in LS~I~+56$\degr$92 showing the strong
single-peaked emission line. Though the shape is typical of a Be star,
the strong 
\ion{C}{ii}~$\lambda\lambda$6578, 6583\AA\ doublet clearly confirms
that this is a star of high luminosity. Continuum units are shown in
the vertical axis.}
\label{fig:alpharara}
\end{figure}

The blue spectrum of LS~I~+56$\degr$92 is shown in
Fig.~\ref{fig:superbe}, together with that of the B3\,II standard
$\iota$ CMa. The degree of Stark broadening in the lines of
LS~I~+56$\degr$92 is at least as low as that in $\iota$ CMa,
confirming that the luminosity class is at least as high. The rather
higher strength of the \ion{Si}{ii}~$\lambda$4128\AA\ doublet compared
to \ion{He}{i}~$\lambda$4121\AA\ and of \ion{Mg}{ii}~$\lambda$4481\AA\
against \ion{He}{i}~$\lambda$4471\AA\ indicate that LS~I~+56$\degr$92
is later than the standard. The weakness of the \ion{Si}{iii} triplet
indicates that this is not a bright supergiant. We conservatively
adopt B4\,II, though B5\,Ib would give a better fit to the distance
modulus of Cam OB3.

Further confirmation of the high luminosity of  LS~I~+56$\degr$92 is
given by the presence of a relatively strong \ion{C}{ii} doublet to
the red of H$\alpha$, as seen in Fig~\ref{fig:alpharara}.

\section{Discussion}

\subsection{Consistency check}

Before attempting to interpret results, it may be sound to check
whether the spectral types derived provide a consistent way of
comparing stars in clusters. For this, I have calculated the
corresponding distance moduli for the stars under study using the
intrinsic colours of Wegner (1994) and the spectral type to intrinsic
magnitude calibration used in Negueruela \& Marco (2003), conveniently
interpolated when necessary. The distances obtained are listed in
Table~\ref{tab:consistent}.

\begin{table*}[ht]
\caption{Distance moduli calculated using the spectral types derived
here and photometry from the literature. The primary reference is
Hiltner (1956), who has been found to be in good accord with other
references when more than one measurement exists. For stars not observed by
Hiltner photometry has been taken from Hoag et al. (1965, BD
$+35\degr$3955) and Coyne et al. (1975, HD~227733).   }
\label{tab:consistent}
\begin{tabular}{llccccc}\hline
Assoc. & Object & Spectral & $V$ & $(B-V)$ & $E(B-V)$& DM\\
& & Type &&&&\\
\hline
Vul OB1 & HD 344863& O9\,III& 8.83 & 0.68 & 0.94 &11.4\\
&HD 344873 & O9.7\,II & 8.77 & 0.77 &1.00 & 11.4\\
\hline
Cyg OB3& HD 227634& B0.2\,II&7.91&0.25&0.46 &12.1\\
&HD 227733& B1.5\,V(e)&10.31 & 0.24&0.46 & 11.7\\
&BD $+35\degr$3955&B0.7\,Iab&7.38 &0.25& 0.45 &12.3\\
&BD $+35\degr$3956&B0.5\,Ve& 8.85 & 0.19&0.43 &11.3\\
\hline
Cyg OB1 &HD 229221 & B0.2\,IIIe& 9.21 & 0.92&1.14 & 10.8\\
&HD 229227 & O9.7\,III & 9.38 & 0.80 & 1.04& 11.6 \\
&HD 229234 & O9\,II& 8.92 & 0.77 &1.04 & 11.6 \\
&HD 229238 & B0.2\,II & 8.88 & 0.90 &1.11 & 11.2\\
&HD 229239 & B0.2\,III & 8.92 & 0.87& 1.10& 10.6 \\
&HD 194280 & OC9.7\,Iab& 8.39 & 0.76&  0.99 & 11.6\\
&HD 194334 & O7\,II(f)& 8.77 & 0.84& 1.13 &11.2\\
\hline
\end{tabular}
\end{table*}

\subsubsection{NGC~6871}

The four stars observed have basically identical $E(B-V)$, fully
compatible with the average $E(B-V)=0.46$ found by previous
workers. Their spectroscopic distances show a large spread, 
but it would fall within what is expected from the method if  BD
$+35\degr$3956 is indeed a binary (for two stars of similar spectral
type, its $DM$ would be $\approx12$). Obviously, none of the Be stars
in this cluster are of high luminosity.

In a recent spectroscopic survey, Balog \& Kenyon (2002) find 6 Be
stars in NGC~6871. All of them appear close to the main sequence. It
appears that no B-type star has moved off the main sequence yet, which
together with the presence of the O-type stars HD~190918 and
HD~190864, favours an age $\la 10$ 
Myr for NGC~6871. This is outside the age range generally quoted for
``Be-rich'' clusters, though the Be fraction appears to be rather
high. 

\subsubsection{Cyg OB1}

The distance moduli have been calculated assuming that $R=3.1$, though
there are strong reasons to believe that the reddening to NGC~6913
does not follow the standard law (Crawford et al. 1977; Wang \& Hu
2000). It is hoped that they can still be used for internal
comparison. 

 Surprisingly, all the luminosity I and II objects give very
consistent distances, but the two objects proposed as Be stars,
which have luminosity class III give rather shorter distances. In the
case of HD~229221, this is partly due to the overestimate of the
interstellar extinction, as the observed $E(B-V)$ must have a
component of circumstellar origin, like in all Be stars. If we rather
assume that all the stars are at the same distance, we find that
HD~229221 and
HD~229239 are (almost) as luminous as cluster members with luminosity
class II, while HD~229227, a fast rotator but not a Be star, is rather
fainter. We have to conclude that the two
confirmed Be stars, HD~229221 and HD~229239 are rather luminous stars.



For Berkeley~87, no attempt has been made at deriving distance moduli,
as none of the stars observed appears to be a single non-emission-line star.

\subsection{Variability}

Variability is a well known characteristic of the Be phenomenon. We
find that several stars that have been classified as emission-line
stars in the past do not display any sign of emission now. In
principle, we will ignore the possibility of a wrong identification
and rather cavalierly assume that the lack of emission lines in modern
spectra must be taken precisely as demonstration of variability,
identifying our targets as Be star (rather than other types of
emission line stars).

The only exceptions to this boldness would be HD~194280 and
HD~194334. The former is well
observed, has a very high luminosity and a single reference to
emission lines. The latter does not look a strong case either.
In the case of BD $+36\degr$4032, we may assume that
the presence of double lines may have induced confusion with emission
components. Moreover, there appears to be no reliable source
identifying the object as an emission-line star.

Among the initial candidates, the three objects in NGC~6871 and
V439~Cyg have been found to be indeed Be
stars, but close to the main sequence, rather than luminous stars.
 We are left with 4 candidates that displayed emission lines at the
time of the observations (HD~229221, HD~207329, LS~I~$+56\degr$92 and 
HD~229059, which may not be a Be star after all), two other candidates
that did not display emission lines, but have been reported as Be
stars by several {\em independent} sources (HD~344863 and HD~229239)
and one possible Be star which has only been reported once
(HD~344873).

Another high-luminosity Be star reported in the literature is
HD~333452. This object was classified 
B0\,III?np by Morgan et al. (1955), the use of both an uncertainty
marker and the ``p'' tag
clearly indicating some peculiarity. The Luminous Star catalogue
classifies it as OB,ce,le,r. However, HD~333452 was observed by Steele et
al.~(1998) in 1998 and appeared as a normal absorption-line O9\,II star.

Another potential high-luminosity Be star is \object{BD
$+56\degr$511}, in \object{$h$ Per}. This object has spectral type
B1\,III (Steele et a. 1999) and is widely reported to be a mild Be star
(cf. Vrancken et al. 2000). Vrancken et al. (2000) find an effective
gravity of only $\log g = 3.1\pm0.1$, based on model atmosphere
analysis, which would suggest a rather high luminosity. However, using
the generally accepted distance modulus to $h$ Per $(m-M)=11.7$ (e.g.,
Marco \& Bernabeu 2001; Keller et al. 2001) and $V=9.11$, $(B-V)=0.38$
from Keller et al. (2001), the implied absolute magnitude (for standard
reddening) is $M_{V}=-4.4$, which does not look particularly high for
B1\,III. 

\subsection{Outlook}
\label{sec:outlook}

As outlined in Section~\ref{sec:intro}, one of the drivers to
undertake this study was the suspicion that stars that appear as B0-1
Be stars could later appear as mid-to-late B-type stars of luminosity
classes I or II, while still keeping their emission lines. It may
therefore appear surprising that all the high-luminosity Be stars
found, with the relative exception of LS~I~$+56\degr$92, are {\em early}
B-type objects.

This might be (at least, in part) due to a selection effect. The
criteria for telling luminosity class II from III are relatively
straightforward for B0 stars, but become more subtle for mid-B stars
and may be rather difficult to apply for emission line objects. It
might then be that some objects classified as Be giants have slightly
higher luminosities but 
have not been recognised as such.
 
The interesting point, however, is the very existence of these
early-type luminous Be
stars. Objects like HD~344863 or HD~333452 must have started their
lives as rather massive stars ($M\ga 25M_{\sun}$), in a region of
the HR diagram (spectral types O8 or earlier) where Be stars are
extremely scarce. As a matter of fact, only one Be star with spectral
type earlier than O8 is well documented in the literature, HD~155806
(O7.5\,IIIe, Conti \& Leep 1974), and even for this one
observations suggest some sort of ``peculiarity'' with respect to other
Be stars (Hanuschik et al. 1996). 
Understanding these very few mid-range O-type stars displaying
emission lines that morphologically resemble Be stars is very important 
in order to
constrain the physical causes leading to the Be phenomenon. Based on
the few examples known, one might suspect that these
objects develop emission lines only late in their lives, but the
crucial point would be to determine whether they actually
bear the same phenomenology as classical Be stars. Therefore this
parameter range will be studied in further work.

Meanwhile, the present work shows that there are at least a few stars
  of high luminosity presenting emission lines that morphologically
  will classify them as Be stars. More detailed analysis of their
  astrophysical parameters is required in order to
  determine whether the resemblance implies a common physical
  mechanism. It would appear, however, rather tempting to conclude that a
  B4II emission-line star displaying the same morphology as a B4IIIe
  star is also a classical Be star.

The number of these luminous Be stars, however, seems to be rather
small, as few convincing candidates were found in the literature and a
substantial fraction do not qualify as such when looked at in
detail. The existence of an upper limit in the luminosity that a Be
star can have appears certain, and the implication that the Be
phenomenon must cease at some point in the life of a Be star has a
bearing on our quest to understand this phenomenon.
 

\acknowledgements
The INT is
operated on the island of La 
Palma by the Isaac Newton Group in the Spanish Observatorio del Roque
de Los Muchachos of the Instituto de Astrof\'{\i}sica de
Canarias. 

This research has made use
of the  Simbad data base, operated at CDS,
Strasbourg, France. 

The author is a researcher of the
programme {\em Ram\'on y Cajal}, funded by the Spanish Ministerio de
Ciencia y Tecnolog\'{\i}a and the University of Alicante.
This research is partially supported by the Spanish Ministerio de
Ciencia y Tecnolog\'{\i}a under grant AYA2002-00814.

The author is very grateful to Nolan Walborn for careful reading of a
first draft and many helpful comments that substantially improved the
paper.

\end{document}